\documentclass[runningheads,a4paper]{llncs}

\usepackage{amssymb}
\usepackage{booktabs}

\usepackage{array}
\usepackage{subfigure}
\usepackage{multirow}

\usepackage{float}
\usepackage{hyperref} 
\usepackage{graphicx}
\usepackage{rotating}
\usepackage{tabularray}
\usepackage{marvosym}
\usepackage{amsmath}
\usepackage{soul}
\usepackage{enumerate}

\usepackage{algorithm}
\usepackage{algpseudocode}

\usepackage{url}

\urldef{\mailsa}\path|{yk.dang, mx.xu, kj.ye}@siat.ac.cn|    
\newcommand{\keywords}[1]{\par\addvspace\baselineskip
\noindent\keywordname\enspace\ignorespaces#1}

\begin{document}

\mainmatter  

\title{Resource Management for GPT-based Model Deployed on Clouds: Challenges, Solutions, and Future Directions}

\titlerunning{Resource Management for GPT-based Model on Clouds}

%
%
\author{Yongkang Dang\and Minxian Xu\and Kejiang Ye}
\authorrunning{Resource Management for GPT-based Model on Clouds}


\institute{Shenzhen Institute of Advanced Technology, Chinese Academy of Sciences\\ Shenzhen, Guangdong, China\\
\mailsa\\
}

%

\toctitle{Lecture Notes in Computer Science}
\tocauthor{Authors' Instructions}
\maketitle 

\begin{abstract}
The widespread adoption of the large language model (LLM), e.g. Generative Pre-trained Transformer (GPT), deployed on cloud computing environment (e.g. Azure) has led to a huge increased demand for resources. This surge in demand poses significant challenges to resource management in clouds. This paper aims to highlight these challenges by first identifying the unique characteristics of resource management for the GPT-based model. Building upon this understanding, we analyze the specific challenges faced by resource management in the context of GPT-based model deployed on clouds, and propose corresponding potential solutions. To facilitate effective resource management, we introduce a comprehensive resource management framework and present resource scheduling algorithms specifically designed for the GPT-based model. Furthermore, we delve into the future directions for resource management in the GPT-based model, highlighting potential areas for further exploration and improvement. Through this study, we aim to provide valuable insights into resource management for GPT-based models deployed in clouds and promote their sustainable development for GPT-based models and applications.
\keywords{GPT-based Model, Cloud Computing, Resource Management, Resource Management Framework, Scheduling Algorithms}
\end{abstract}

\section{Introduction}
\begin{figure}[ht]               
\centering
\includegraphics[width=\textwidth]{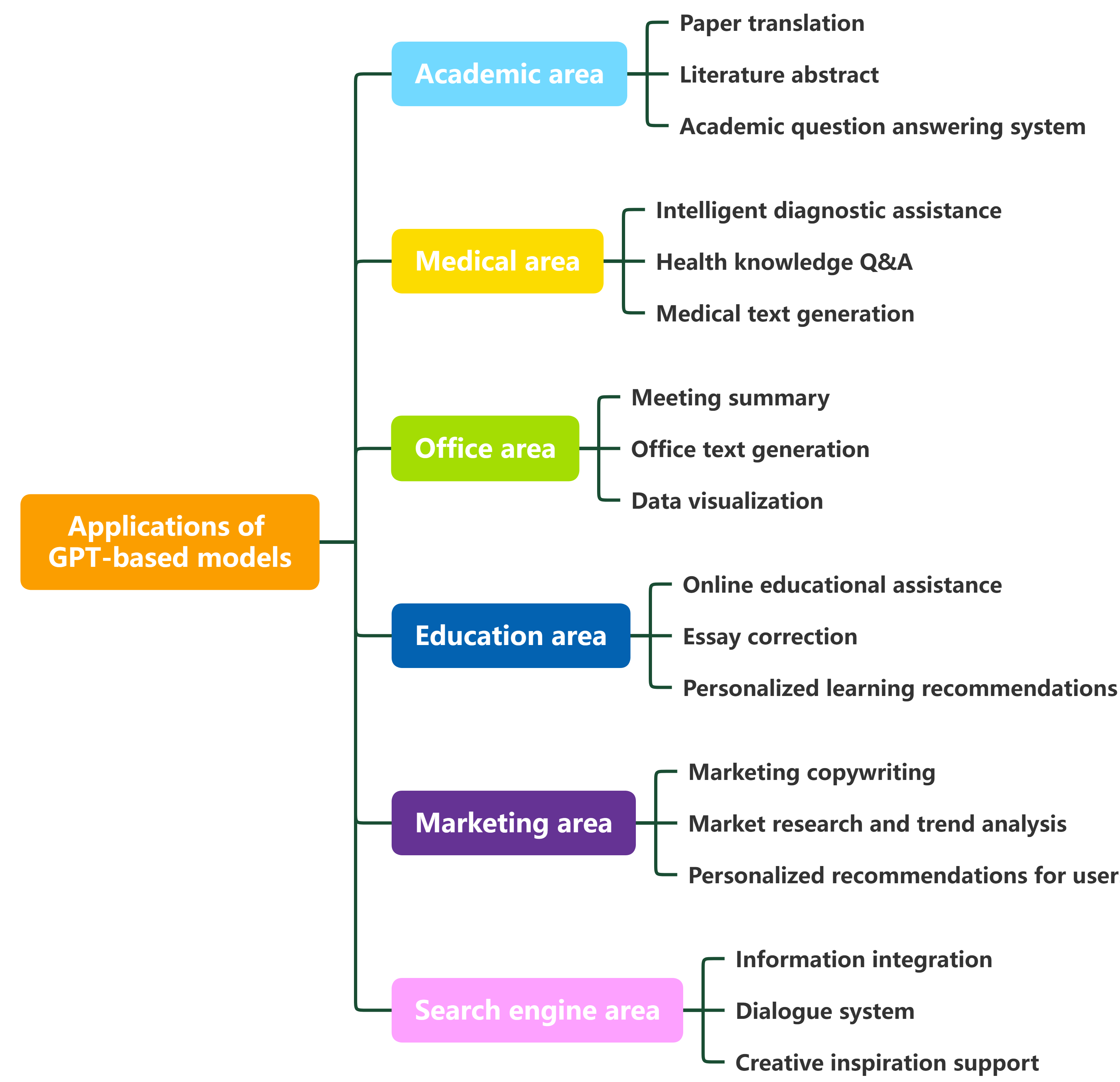}
\caption{GPT-based Model Application Areas}
\label{GPT-based Model Application Areas}
\end{figure}

The GPT-based model is a language generation model based on the transformer architecture, which learns the statistical regularities and semantic knowledge of language through unsupervised pre-training on large-scale text datasets. The model can then be fine-tuned on specific domain or task data through supervised or semi-supervised learning to adapt to different language application scenarios \cite{radford2018improving} \cite{brown2020language}. The GPT-based model can generate natural, fluent, context-aware, and semantically coherent language content, making it suitable for applications such as text summarization, machine translation, sentiment analysis, question answering systems, and chatbots. Figure~\ref{GPT-based Model Application Areas} provides examples of current applications of the GPT-based model for different areas\cite{nori2023capabilities} \cite{bommarito2023gpt}, including academic, medical, office, education \cite{GILL202419} and marketing. 

The GPT-based model was first proposed and developed by OpenAI and has since been developed into multiple versions and variants, such as GPT-1, GPT-2, GPT-3, and GPT-4. The main differences between these models lie in the number of parameters, dataset size, and training methods. For example, GPT-3 is currently one of the largest GPT-based models, with 175 billion parameters, requiring 800 GB of storage \cite{radford2018improving}. It has achieved excellent performance on multiple natural language processing tasks. Table~\ref{Parameters of Mainstream GPT-based Models} lists several mainstream GPT-based models, providing relevant details such as model names, research teams, release dates, and model sizes, with each row representing a distinct model.

\begin{table}[ht]
    \centering
    \caption{Parameters of Mainstream GPT-based Models \cite{zhao2023survey}}
    \begin{tabular}{cccc}
    \toprule
        Model & Research Teams & Release Date & Size \\ \midrule
        GPT-3 & OpenAI & 28-May-20 & 175B \\ 
        PanGu-$\alpha$ & Pengcheng Laboratory and other teams & 26-Apr-21 & 207B \\ 
        OPT-175B & Meta & 3-May-22 & 175B \\ 
        PaLM &  Google & Apr-22 & 540B \\ 
        BLOOM & BigScience & Dec-22 & 176B \\ 
        MT-NLG & Microsoft and NVIDIA & 11-Oct-21 & 530B \\ 
        Gopher & DeepMind & 8-Dec-21 & 280B \\ 
        CPM-2 & Beijing Academy of Artificial Intelligence & Jun-21 & 198B \\ 
        GPT-Neo-X-20B & EleutherAI  & Apr-22 & 20B \\
        \bottomrule
    \end{tabular}
    \label{Parameters of Mainstream GPT-based Models}
\end{table}

The rapid development and widespread application of GPT-based model have led to the increased demand for resources, and the current GPT-based model has been deployed on public clouds like Azure and Google Cloud for training and inference, as GPT-based models are typically too large and resource-intensive to be deployed on edge devices or small-scale hardware. Therefore, they are better suited for cloud-based deployments,
which also makes the resource management for GPT-based model facing some specific challenges.
 In order to summarize these challenges and propose corresponding solutions, in this section, we will identify the unique characteristics of resource management for GPT-based models and establish evaluation metrics for this specific domain.

\subsection{Unique Characteristics of Resource Management for GPT-based Model}

Through extensive research, we identified the following unique characteristics of resource management for GPT-based model:

\textbf{Large-scale computational demands due to huge amount of parameters and fine-tuning:} The GPT-based model typically consists of billions of parameters, necessitating a substantial amount of computational resources during both training and inference processes \cite{fedus2022switch}. Training GPT models typically requires specialized hardware such as Graphics Processing Units (GPUs) or Tensor Processing Units (TPUs) due to the sheer number of calculations involved. In addition, fine-tuning a pre-trained GPT model on a specific task requires additional compute resources, as the model needs to adapt to the task through further training.
This complexity makes resource management more intricate, requiring efficient allocation and utilization of computational resources to ensure the model operates efficiently.

\textbf{High storage demands to support rapid data access:} The GPT-based model's large parameter size requires significant storage space to accommodate model parameters and intermediate computation results. Running these models can quickly consume all available memory on conventional hardware. Therefore, resource management must consider how to effectively manage storage resources to meet the model's requirements while ensuring rapid data access and processing.

\textbf{High-speed network demands to enable efficient parallelism:} During model training, the GPT-based model handles vast datasets and performs complex computations and parameter optimization, demanding fast data transmission and stable network connectivity. In the inference phase, efficient network resource utilization directly impacts inference speed and response time. The GPT-based model needs to generate outputs based on inputs and provide results in real-time or near real-time conditions. Hence, network resources are crucial for achieving fast responses and efficient inference.

\textbf{Long training and inference processes than traditional AI models:} Traditional AI applications often have lower computational requirements and faster inference times. However, due to the complexity and scale of the GPT-based model, its training and inference processes typically require extended periods. Resource management must consider how to maintain system stability and performance over an extended duration while ensuring the rational allocation and utilization of resources.

\textbf{Dynamic resource demands from varied complexity of tasks:} In practical applications, the resource requirements of the GPT-based model may vary over time and across different tasks, e.g. machine translation, text summarization and question answering. These tasks can have dynamic resources demand due to the different degree of complexity (e.g. difficulty of questions and expected output length). 
Resource management must possess dynamic adjustment capabilities, allowing for the dynamic allocation of computational and storage resources based on actual demands to adapt to different stages and tasks.

By understanding and addressing these unique characteristics, effective resource management strategies can be developed to ensure optimal performance and utilization of  the GPT-based model in various applications.

\subsection{Evaluation Metrics for Resource Management for GPT-based Model}

In order to effectively evaluate the resource management of GPT-based model, we can consider the following metrics:

\textbf{Resource Utilization:} It refers to the degree to which the model effectively utilizes available resources during the training or inference process. For the GPT-based model, resources primarily include computational resources (such as CPUs and GPUs), storage resources (such as memory and disk space), and network resources. Evaluating resource utilization involves ensuring that the model maximizes the use of available resources to improve efficiency and minimize resource waste. This can be achieved through optimization of scheduling algorithms and parallel computing techniques. Higher resource utilization indicates efficient utilization of computational, storage, and network resources, enhancing overall system performance.

\textbf{Time Efficiency:} It refers to the time taken by the model to complete a set of  given tasks (e.g. makespan metric used in traditional task scheduling). For the GPT-based models, time efficiency includes both model training time and inference time. During training, time efficiency focuses on the speed of parameter updates on a given dataset. Training time efficiency can be improved through optimization of scheduling algorithms, distributed training, and hardware acceleration. During inference, time efficiency concerns the speed at which the model processes input data and generates outputs. Inference time directly affects the real-time performance and responsiveness of the model in practical applications. Techniques such as parallel computing, batch processing, and hardware optimization can improve inference time efficiency. Higher time efficiency means the model can complete training and inference tasks more quickly, thereby increasing overall production efficiency.

\textbf{Cost Efficiency:} The cost of  the GPT-based model mainly includes computational cost, storage cost, and network transmission cost. Computational cost evaluation primarily considers the computational resource expenses required by the model. This includes model complexity, computational workload, and the hardware used. Lower computational costs imply relatively lower computational resource expenses required for specific tasks, reducing resource investments. Storage cost refers to the expenses associated with storage resources needed for the model, including model parameters, intermediate results, and cached data. Storage cost can be measured based on the model's size and the capacity of the storage devices used. Lower storage costs indicate that the model occupies a relatively smaller storage space, reducing storage resource demands and related expenses. Network transmission cost involves the expenses associated with network resources for data transmission during the model training or inference process, including model parameter transmission, training data transmission, and inference result transmission. Lower network transmission costs mean the model efficiently utilizes network resources, reducing data transmission time and bandwidth expenses.

This paper aims to highlight the specific challenges in resource management for GPT-based models and propose corresponding solutions. The main contributions of this paper are as follows:

\textbullet\ we summarize the specific challenges in resource management for GPT-based model and provide a detailed description of these challenges.

\textbullet\ We propose a comprehensive resource management framework for the GPT-based model that comprises seven different functional components.

\textbullet\ we present three resource management algorithms specifically designed for the GPT-based model to optimize the resource usage based on different objectives including resource utilization, load balancing and energy efficiency.

The rest of this paper is organized as follows: In Section \ref{sec: Specific Challenges in Resource Management for GPT-based Model}, we will summarize the specific challenges in resource management for GPT-based model. By highlighting these challenges, we aim to provide directions for corresponding solutions. To facilitate effective resource management, in Section \ref{sec: Resource Management Framework for GPT-based Model}, we will propose a comprehensive resource management framework for the GPT-based model. Additionally, in Section \ref{sec: Resource Management Algorithms for GPT-based Model}, we will introduce three resource management algorithms specifically tailored to the GPT-based model. Finally, we will conclude this work and discuss several future research directions for resource management for GPT-based model in Section \ref{sec: Future Research Directions}.

\section{Specific Challenges in Resource Management for GPT-based Model}\label{sec: Specific Challenges in Resource Management for GPT-based Model}

\begin{figure}[ht]               
\centering
\includegraphics[width=\textwidth]{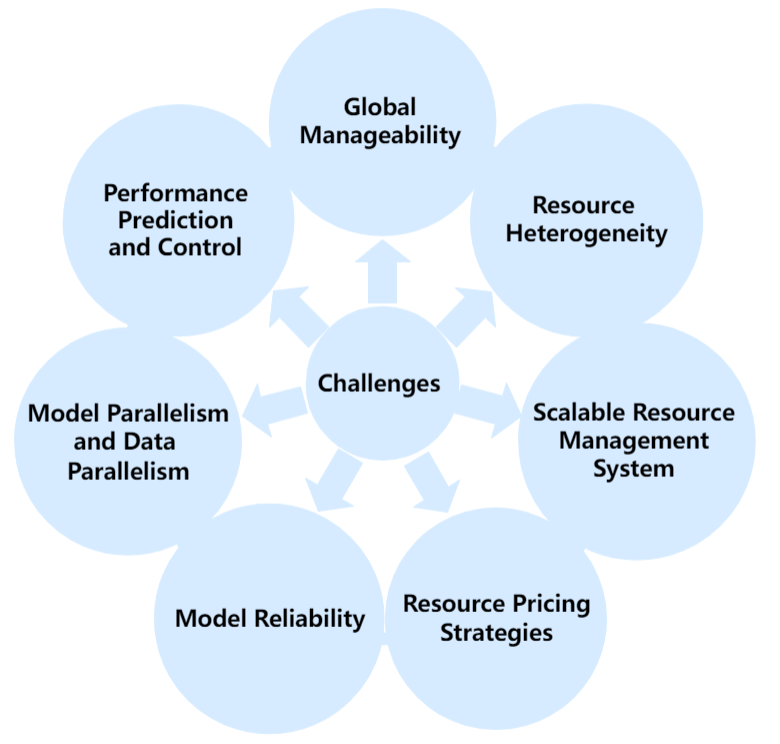}
\caption{Challenges in Resource Management for GPT-based Model}
\label{Specific Challenges in Resource Management for GPT-based Model}
\end{figure}

By identifying the unique characteristics of resource management for GPT-based model, we summarized the specific challenges in resource management for GPT-based model deployed on clouds \cite{jennings2015resource} \cite{xu2022esdnn} \cite{cortez2017resource} \cite{gunasekaran2020implications} \cite{zhang2019resource} \cite{li2023alpaserve}. For a more visual representation, we presented these specific challenges in Figure~\ref{Specific Challenges in Resource Management for GPT-based Model}. We noted some main challenges such as performance prediction and control, global manageability, resource heterogeneity, scalable resource management system, resource pricing strategies, model reliability, model parallelism and data parallelism, and we will discuss the details in the following sections.

\subsection{Performance Prediction and Control}

The GPT-based model, constructed based on the Transformer architecture, has a large parameter size and complex structure. This complexity leads to significant computational requirements and the need for a large number of parameters for training and inference. Different tasks typically have varying complexities, resource demands, and performance expectations for the model. Even under the same workload and resource configuration, the performance of the model can be influenced by task characteristics and data properties. Additionally, different workloads and resource configurations can lead to variations in resource allocation, data parallelism, and other aspects that influence performance. These factors make it challenging to predict and control the behavior and performance of the model under different workloads and resource configurations.

\subsection{Global Manageability}

Global manageability refers to effectively managing and coordinating resources, including computational resources, storage resources, and network resources, in large and complex cloud environments. In the context of GPT-based model applications, challenges in achieving global manageability primarily manifest in the following aspects:

\textbf{Resource scheduling and allocation:} Given the massive computational, storage, and network resource requirements of the GPT-based model, efficient resource scheduling and allocation algorithms are needed. This includes dynamic resource allocation across different data centers and geographical locations to meet user demands and service-level agreements.

\textbf{Resource monitoring and optimization:} Achieving global manageability requires real-time monitoring of resource usage, performance metrics, and health status, coupled with automated techniques for resource adjustments and optimization. Such monitoring mechanisms help maintain efficient resource utilization, ensure load balancing, and optimize performance bottlenecks, thereby enhancing overall system performance.

\subsection{Resource Heterogeneity}

Resource heterogeneity refers to the existence of various types or characteristics of resources within the same system or environment. These resources can include computational resources (such as CPUs, GPUs, and TPUs), storage resources (such as disks and solid-state drives), network resources (such as bandwidth and latency), and others. Resource heterogeneity implies differences in performance, scale, power consumption, and cost among these resources. For the GPT-based model, resource heterogeneity poses the following challenges:

\textbf{Resource dependencies:} Resource dependencies refer to the interdependencies and associations among different types of resources. In resource management, it is necessary to consider these dependencies and employ suitable resource allocation algorithms to optimize the synergistic effects among resources, thereby improving overall system performance. For example, in the GPT-based model, the supply of computational resources must match the capacity of storage resources and network bandwidth to ensure efficient data transmission and smooth model operation. By fully considering resource dependencies, resource allocation and utilization can be optimized, maximizing the system's potential.

\textbf{Resource interoperability:} Different types of resources are often provided by different vendors and technologies, necessitating addressing the challenge of resource interoperability. This involves establishing standards and protocols to ensure seamless integration and interaction among different types of resources, improving system compatibility and interoperability. Additionally, data and model transfer and sharing across different resources need to be addressed to enable collaborative work across resources.

\subsection{Scalable Resource Management System}

With the development of the GPT-based model, its scale has increased significantly, demanding huge computational resources. Moreover, as the GPT-based model is widely applied, data centers face concurrent requests and high throughput demands. Therefore, a highly scalable computing and storage infrastructure is required to support model execution and handle massive data. Furthermore, the resource management system must scale across multiple dimensions to handle resource management in large-scale data centers. The GPT-based model requires efficient management and allocation of computational, storage, and network resources, as well as task scheduling, to meet the training and inference requirements of the model. Additionally, the resource management system needs to dynamically allocate and flexibly expand resources to accommodate different application scenarios with varying scales and complexities. These requirements pose important challenges to the resource management system for the GPT-based model.

\subsection{Resource Pricing Strategies}

Resource pricing strategies are crucial for the GPT-based model as they directly impact resource utilization, user satisfaction, and vendor profitability. However, several challenges exist in resource pricing strategies.

Firstly, accurately determining resource costs is a challenge. Resource costs are influenced by factors such as the vendor, geographical location, and usage volume. Therefore, a comprehensive consideration of these factors is required to ensure that resource pricing covers actual costs and attracts user adoption.

Secondly, balancing the supply-demand relationship is another challenge. Vendors aim to obtain revenues through resource sales while ensuring stable supply, while users seek resources at reasonable prices and sufficient support during peak demand. Therefore, resource pricing strategies need to strike a balance between supply and demand, meeting user needs while ensuring vendor profitability.

Additionally, achieving fair resource allocation and pricing is also a challenge. In multi-user environments, resources must be allocated on-demand to different users and priced based on usage. Given the variations in user demands and usage patterns, assuring fair resource allocation and pricing becomes a complicated problem.

\subsection{Model Reliability}

Due to the complexity of the GPT-based model, such as its large scale and long training processes, model failures during operation are inevitable. To ensure model reliability, systems must implement fault detection and fault tolerance mechanisms to handle resource failures or interruptions promptly. Fault detection mechanisms proactively identify potential system failures by monitoring model performance metrics, resource utilization, and other key parameters. Fault tolerance mechanisms include data backup and recovery strategies to ensure data integrity and service continuity. Data backup strategies involve regular backups of the GPT-based model parameters, training data, and other related data to ensure available backup data for recovery in case of failures. Recovery strategies ensure quick system recovery and maintain the continuity of user experience after failures or interruptions. Through these mechanisms, the system can rapidly detect and respond to faults, reducing the risks of system downtime and data loss, thus ensuring the reliability of the GPT-based model.

\subsection{Model Parallelism and Data Parallelism}

In model parallelism, challenges primarily include:

\textbf{Model partitioning:} The GPT-based model typically has a large scale, consisting of billions or even hundreds of billions of parameters. Partitioning such a massive GPT-based model into sub-models suitable for parallel processing is a challenge. Model partitioning needs to consider the dependencies within the model structure and the communication requirements among parameters to ensure correctness and efficiency in parallel computation.

\textbf{Synchronization and communication overhead:} Synchronization and communication are required among sub-models on different devices to ensure the transfer and aggregation of gradient information during training and enable effective parameter updates. Synchronization and communication operations can introduce additional computational and communication overhead, impacting the efficiency and performance of parallel computation.

\textbf{Load balancing:} Proper distribution of computational load is crucial for model parallelism to ensure balanced computation across devices. Load imbalance can result in computational resource waste and decreased efficiency.

In data parallelism, challenges primarily include:

\textbf{Data partitioning and distribution:} The GPT-based model has massive training data, and partitioning the data into multiple parts and distributing them to different devices is a challenge. Data partitioning needs to consider data balance and distribution efficiency to ensure the quality and performance of parallel training.

\textbf{Data synchronization and consistency:} In data parallel computation, model synchronization and consistency are crucial to ensure accurate parameter updates. Efficient data synchronization mechanisms are key to ensuring the effectiveness of parallel training.

\textbf{Training speed limitations:} In data parallel computation, the training speed may be limited by the slowest device. If some devices have slower computation speeds, it will affect the overall training efficiency and speed.

\section{Resource Management Framework for GPT-based Model}\label{sec: Resource Management Framework for GPT-based Model}

\begin{figure}[ht]               
\centering
\includegraphics[width=0.8\textwidth]{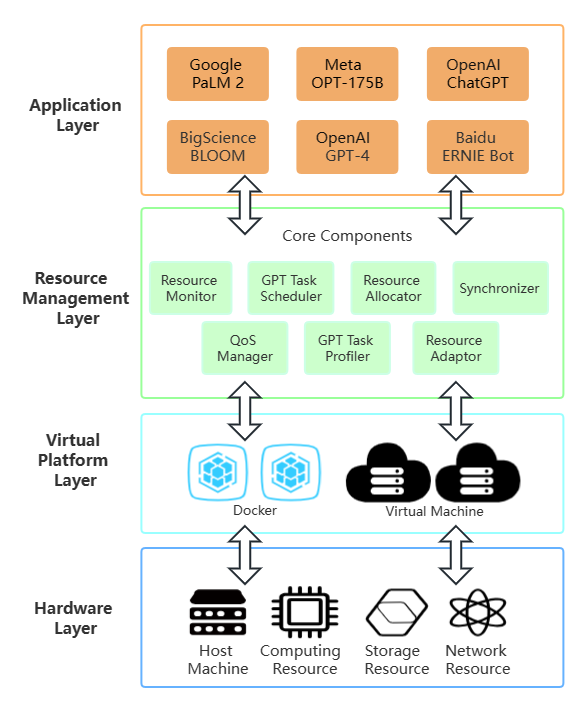}
\caption{Resource Management Framework for GPT-based Model}
\label{Resource Management Framework for GPT-based Model}
\end{figure}

In response to the specific challenges faced by the GPT-based model and based on the characteristics of  resource management for GPT-based model, we propose a comprehensive resource management framework. This framework aims to effectively manage critical resources such as computational resources, storage resources, and network bandwidth required by the GPT-based model, thereby improving overall model efficiency and ensuring model reliability and service quality. Figure~\ref{Resource Management Framework for GPT-based Model} demonstrates our resource management framework for the GPT-based model. The resource management framework is divided into several key components, including Resource Monitor, GPT Task Scheduler, Resource Allocator, GPT Task Profiler, Synchronizer, QoS Manager, and Resource Adaptor. The following will provide detailed introductions for each of these components. 

\textbf{Resource Monitor:} The Resource Monitor is responsible for real-time monitoring of computational resources (e.g. CPU, GPU, memory), network resources (e.g. bandwidth utilization, network latency), and storage resources (e.g. disk) in the system. It collects and analyzes real-time resource usage and performance data, providing real-time feedback and reports to support task scheduling and resource allocation decisions. It also offers visual representations to show resource usage and performance metric trends.

\textbf{GPT Task Scheduler:} The GPT Task Scheduler handles task scheduling based on requests from the GPT task queue. It considers task priority, resource requirements, timeliness, and related constraints to select suitable GPT-based model instances for task scheduling. By employing appropriate scheduling algorithms, it determines the execution order of tasks and assigns them to available GPT model instances.

\textbf{Resource Allocator:} The Resource Allocator dynamically manages system resources based on task resource requirements, system resource availability, and load conditions. It employs intelligent resource allocation strategies to meet the execution needs of tasks. Additionally, the Resource Allocator may utilize resource prediction and load forecasting models to predict task resource requirements and system load conditions, enabling more accurate resource allocation and adjustments.

\textbf{GPT Task Profiler:} The GPT Task Profiler extracts attributes and requirements of GPT tasks for better understanding and handling. The functionalities of the task profiler include:

1) Task Attribute Extraction: The GPT Task Profiler extracts task types, input data features (e.g. text length, language style, domain-specific vocabulary), and requirements for answering. These attributes contribute to better task understanding, enabling customized parameter settings for corresponding GPT model instances and providing essential references for subsequent data processing and answer generation.

2) Model Resource Requirement Analysis: The GPT Task Profiler analyzes the resource requirements of GPT tasks, including computational resources and storage resources. By evaluating the resource requirements of GPT tasks, the GPT Task Profiler provides accurate foundations for resource allocation and scheduling.

3) QoS Requirement Extraction: The GPT Task Profiler identifies QoS requirements of GPT tasks, such as response time, text generation speed, and answer accuracy. It transfers these requirements to the QoS Manager for evaluation and management of task QoS requirements.

\textbf{Synchronizer:} The Synchronizer ensures reasonable resource allocation and smooth execution of GPT tasks by employing distributed consistency protocols. The Synchronizer includes the following functionalities:

1) Distributed Lock Management: The Synchronizer uses distributed lock mechanisms to prevent conflicts and competitions for shared resources. This ensures that only one thread or process can gain exclusive access to resources at a time.

2) Task State Synchronization: The Synchronizer ensures consistent task states across different nodes through distributed consistency protocols. This includes task states such as start, pause, terminate, and completion. By efficient negotiation and communication, the Synchronizer promptly updates and shares task state information, enabling other components to accurately understand task execution progress.

3) Transaction Handling: The Synchronizer employs distributed consistency protocols for transactional processing of resource management operations. This means that resource allocation, release, and scheduling operations are atomic and maintain system consistency. Even in the event of failures or anomalies, the Synchronizer can perform rollbacks or recovery operations, ensuring the reliability and correctness of resource management.

4) Fault Recovery: The Synchronizer handles node failures and network anomalies through distributed consistency protocols. It detects node failures and takes corresponding measures, such as re-election or reallocation of resources, to maintain system availability and stability.

\textbf{QoS Manager:} The QoS Manager evaluates and manages the Quality of Service (QoS) requirements of GPT tasks. The QoS Manager includes the following functionalities:

1) QoS Requirement Evaluation: The QoS Manager receives GPT task QoS requirement information from the GPT Task Profiler, such as response time, processing time, and answer accuracy requirements. It evaluates and quantifies these requirements to understand different task QoS demands.

2) QoS Optimization Strategy Formulation: Based on the QoS requirements of GPT tasks and constraints of system resources, the QoS Manager formulates corresponding optimization strategies. It considers the urgency of GPT tasks and QoS requirements, combined with system resource availability and constraints, to provide guidance and suggestions for the GPT Task Scheduler and Resource Allocator. By adjusting resource allocation priorities, scheduling algorithm parameters, and other strategies, the QoS Manager ensures that QoS requirements of GPT tasks are met.

3) QoS Monitoring and Feedback: The QoS Manager monitors the real-time performance of GPT tasks and QoS metrics. It regularly checks QoS indicators, such as response time, processing time, and accuracy, and compares them with GPT task QoS requirements. If the QoS requirements are not met, the QoS Manager initiates appropriate resource adjustments and optimizations.

\textbf{Resource Adaptor:} The Resource Adaptor is responsible for dynamic resource scaling based on system load, GPT task demands, and resource usage. It uses adaptive algorithms and prediction models to automatically adjust resource allocation for GPT tasks, achieving dynamic resource expansion and contraction. When the system load or task demands increase, the Resource Adaptor automatically scales up resources to meet the requirements and maintain system performance. Conversely, during low system load or reduced task demands, it timely scales down resources to reduce waste. By optimizing resource allocation through the Resource Adaptor, resource distribution becomes more flexible and intelligent, capable of adapting to dynamically changing resource demands.

\section{Resource Management Algorithms for GPT-based Model}\label{sec: Resource Management Algorithms for GPT-based Model}

Based on our resource management framework for GPT-based model, we propose three resource management algorithms: Maximization of Compute Resource Utilization, Load Balancing, and Power-efficient optimization, as the resource utilization, load balancing efficiency and power efficiency are the most dominant concerns of resource scheduling approaches \cite{xu2017survey} \cite{tuli2022hunter} \cite{beloglazov2012optimal} \cite{beloglazov2012energy}. These algorithms aim to allocate available nodes for user requests to achieve efficient resource management. Through these algorithms, we can maximize the utilization of computing resources, achieve load balancing, and optimize energy consumption. The table~\ref{Variable Definitions} provides the definitions of variables used in the algorithms.

\begin{table}[ht]
\centering
\caption{Variable Definitions}
\begin{tabular}{lp{8cm}}
\toprule
\textbf{Variable} & \textbf{Definition} \\
\midrule
\texttt{gptRequestQueue} & A queue containing GPT model requests that need to be processed. \\
\texttt{availableNodes} & A list of available nodes that can be used to process the GPT model requests. \\
\texttt{threshold} & A utilization threshold (0 to 1) that determines the maximum utilization allowed for each resource on a node. \\
\texttt{allocationMap} & A dictionary that maps each GPT model request to an allocated node. \\
\texttt{request} & The current GPT model request being processed. \\
\texttt{node} & The current node being processed. \\
\texttt{computeDemand} & The required compute resource for a request. \\
\texttt{memoryDemand} & The required memory resource for a request. \\
\texttt{storageDemand} & The required storage resource for a request. \\
\texttt{computeCapacity} & The compute resource capacity of a node. \\
\texttt{memoryCapacity} & The memory resource capacity of a node. \\
\texttt{storageCapacity} & The storage resource capacity of a node. \\
\texttt{computeDemandPercentage} & The percentage of compute resource requested by the current request, relative to the compute resource capacity of the current node. \\
\texttt{memoryDemandPercentage} & The percentage of memory resource requested by the current request, relative to the memory resource capacity of the current node. \\
\texttt{storageDemandPercentage} & The percentage of storage resource requested by the current request, relative to the storage resource capacity of the current node. \\
\texttt{computeUtilization} & The compute resource utilization of a node. \\
\texttt{memoryUtilization} & The memory resource utilization of a node. \\
\texttt{storageUtilization} & The storage resource utilization of a node. \\
\texttt{allocated} & A flag indicating whether a request has been allocated to a node or not. \\
\texttt{minPower} & The minimum power value among all the scanned nodes. \\
\bottomrule
\end{tabular}
\label{Variable Definitions}
\end{table}

\subsection{Maximization of Compute Resource Utilization}
The goal of algorithm \ref{Maximization of Compute Resource Utilization} is to optimize the task-to-node assignment by maximizing the utilization of computing resources. Taking into account the threshold of node resources, the algorithm prioritizes nodes with high resource utilization to serve the requests, thereby avoiding resource waste and maximizing the utilization of node resources. The optimization objective equation for the algorithm is given in (\ref{equation 1}).

\begin{equation}
\label{equation 1}
\max \frac{\sum\limits_{i=1}^k u(N_i)}{k},
\end{equation}
where  $u(N_i)$ denotes the compute resource utilization of node $N_i$, $\sum\limits_{i=1}^k$ denotes the summation over the compute resource utilization of each node, and $k$ denotes the number of available nodes.

Algorithm \ref{Maximization of Compute Resource Utilization} takes as input a queue of GPT model requests, a list of available nodes, and a threshold value that determines the maximum utilization allowed for each resource on a node. The algorithm aims to maximize the utilization of compute resources while ensuring that the utilization of each resource on a node does not exceed the specified threshold.

To achieve this, the algorithm first estimates the resource demand of each request in the queue by calling the function request.estimateResourceDemand() for each request (line 4). It then sorts the queue in descending order of compute resource demand (line 6) and the available nodes in descending order of compute utilization (line 7).

For each request in the queue, the algorithm attempts to allocate a node that can satisfy the request's resource demands without exceeding the threshold for any resource. It does this by iterating over each available node and computing the percentage of compute, memory, and storage resources that would be consumed by the request if it were allocated to that node (lines 11-14). If the node's compute, memory, and storage utilizations, plus the percentage of resources demanded by the request, are all less than or equal to the threshold, the node is allocated to the request (lines 15-20).

If no available node can satisfy the request's resource demands, the algorithm creates a new node, allocates it to the request, and updates its utilization accordingly (lines 25-30).

The algorithm returns a dictionary, allocationMap, where each key is a GPT request and each value is the node allocated to that request.

\begin{algorithm}[!htbp]
\caption{Maximization of Compute Resource Utilization for GPT requests}
\label{Maximization of Compute Resource Utilization}
\begin{algorithmic}[1]
\State \textbf{Input}\textbf{: }gptRequestQueue, availableNodes, threshold
\State \textbf{Output:} allocationMap

\For{\textbf{each} request \textbf{in} gptRequestQueue}
    \State request.estimateResourceDemand() \Comment{Estimate resource demand for each request}
\EndFor
\State gptRequestQueue.sortByDescendingComputeResourceDemand() \Comment{Sort requests by descending compute resource demand}
\State availableNodes.sortByDescendingComputeUtilization() \Comment{Sort available nodes by descending compute utilization}
\State $allocationMap \gets \{\}$

\For{\textbf{each} request \textbf{in} gptRequestQueue}
    \State allocated $\gets$ \textbf{false}
    \For{\textbf{each} node \textbf{in} availableNodes}
        \State computeDemandPercentage $\gets$ request.computeResourceDemand / node.computeCapacity
        \State memoryDemandPercentage $\gets$ request.memoryResourceDemand / node.memoryCapacity
        \State storageDemandPercentage $\gets$ request.storageResourceDemand / node.storageCapacity

        \If{(node.computeUtilization + computeDemandPercentage $\leq$ threshold) and (node.memoryUtilization + memoryDemandPercentage $\leq$  threshold) and (node.storageUtilization + storageDemandPercentage $\leq$ threshold)}
            \State allocationMap[request] $\gets$ node \Comment{Allocate request to the current node}
            \State node.computeUtilization += computeDemandPercentage \Comment{Update node's compute utilization}
            \State node.memoryUtilization += memoryDemandPercentage \Comment{Update node's memory utilization}
            \State node.storageUtilization += storageDemandPercentage \Comment{Update node's storage utilization}
            \State allocated $\gets$ \textbf{true}
            \State \textbf{break}
        \EndIf
    \EndFor
    \If{\textbf{not} allocated}
        \State newNode $\gets$ createNewNode() \Comment{Create a new node}
        \State availableNodes.append(newNode) \Comment{Add the new node to the list of available nodes}
        \State allocationMap[request] $\gets$ newNode \Comment{Allocate request to the new node}
        \State newNode.computeUtilization += computeDemandPercentage 
        \State newNode.memoryUtilization += memoryDemandPercentage 
        \State newNode.storageUtilization += storageDemandPercentage 
        \State allocated $\gets$ \textbf{true}
    \EndIf
\EndFor
\State \textbf{return} allocationMap \Comment{Return the allocation mapping}
\end{algorithmic}
\end{algorithm}

\subsection{Load Balancing}

The purpose of algorithm \ref{Load Balancing} is to optimize the selection of tasks to nodes by balancing the load across nodes. Considering the thresholds of node resources, this algorithm prioritizes serving requests from nodes with lower resource utilization, thereby ensuring load balancing in the system. The optimization objective equation for the algorithm is given in (\ref{equation 2}).

\begin{equation}
\label{equation 2}
\min \sqrt{\frac{1}{k} \sum_{i=1}^{k} (u(N_i) - \bar{u})^2},
\end{equation}
where $k$ denotes the number of available nodes, $u(N_i)$ denotes the compute resource utilization of node $N_i$, and $\bar{u}$ denotes the average compute resource utilization across all available nodes.

Algorithm \ref{Load Balancing} is similar to Algorithm 1, but the difference lies in sorting the list of available nodes in ascending order of compute resource utilization. This prioritizes the allocation to nodes with lower utilization rates. Other steps remain the same as Algorithm 1.

\begin{algorithm}[!htbp]
\caption{Load Balancing Algorithm for GPT requests}
\label{Load Balancing}
\begin{algorithmic}[1]
\State \textbf{Input}\textbf{: }gptRequestQueue, availableNodes, threshold
\State \textbf{Output:} allocationMap

\For{\textbf{each} request \textbf{in} gptRequestQueue}
    \State request.estimateResourceDemand() \Comment{Estimate resource demand for each request}
\EndFor
\State gptRequestQueue.sortByDescendingComputeResourceDemand() \Comment{Sort requests by descending compute resource demand}
\State availableNodes.sortByAscendingComputeUtilization() \Comment{Sort available nodes by ascending compute utilization}
\State $allocationMap \gets \{\}$

\For{\textbf{each} request \textbf{in} gptRequestQueue}
    \State allocated $\gets$ \textbf{false}
    \For{\textbf{each} node \textbf{in} availableNodes}
        \State computeDemandPercentage $\gets$ request.computeResourceDemand / node.computeCapacity
        \State memoryDemandPercentage $\gets$ request.memoryResourceDemand / node.memoryCapacity
        \State storageDemandPercentage $\gets$ request.storageResourceDemand / node.storageCapacity

        \If{(node.computeUtilization + computeDemandPercentage $\leq$ threshold) and (node.memoryUtilization + memoryDemandPercentage $\leq$  threshold) and (node.storageUtilization + storageDemandPercentage $\leq$ threshold)}
            \State allocationMap[request] $\gets$ node \Comment{Allocate request to the current node}
            \State node.computeUtilization += computeDemandPercentage \Comment{Update node's compute utilization}
            \State node.memoryUtilization += memoryDemandPercentage \Comment{Update node's memory utilization}
            \State node.storageUtilization += storageDemandPercentage \Comment{Update node's storage utilization}
            \State allocated $\gets$ \textbf{true}
            \State \textbf{break}
        \EndIf
    \EndFor
    \If{\textbf{not} allocated}
        \State newNode $\gets$ createNewNode() \Comment{Create a new node}
        \State availableNodes.append(newNode) \Comment{Add the new node to the list of available nodes}
        \State allocationMap[request] $\gets$ newNode \Comment{Allocate request to the new node}
        \State newNode.computeUtilization += computeDemandPercentage 
        \State newNode.memoryUtilization += memoryDemandPercentage 
        \State newNode.storageUtilization += storageDemandPercentage 
        \State allocated $\gets$ \textbf{true}
    \EndIf
\EndFor
\State \textbf{return} allocationMap \Comment{Return the allocation mapping}
\end{algorithmic}
\end{algorithm}

\subsection{Power-efficient Optimization}

The purpose of algorithm \ref{Power-efficient Optimization} is to optimize the selection of tasks to nodes by minimizing energy consumption, aiming to minimize the total energy consumption of available nodes. This algorithm prioritizes serving requests on nodes with the lowest energy consumption, thus achieving power-efficient optimization. The optimization objective formula for the algorithm is given in (\ref{equation 3}).

\begin{equation}
\label{equation 3}
\text{min } \sum_{i=1}^{k} \text{power}(N_i),
\end{equation}
where $k$ denotes the number of available nodes and $\text{power}(N_i)$ denotes the power consumption of node $N_i$.

Algorithm \ref{Power-efficient Optimization} requires two input parameters: the request queue gptRequestQueue and the list of available nodes availableNodes (Line 1). The output is the allocation mapping allocationMap (Line 2).

Firstly, the algorithm creates an empty allocation mapping allocationMap (Line 3), and iterates through each request in the request queue, using the estimateResourceDemand() function to estimate its resource demand (Line 4-5). Then, the algorithm sorts the request queue in descending order based on the resource demand, to handle requests that require more resources first (Line 7). Next, the algorithm iterates through each request in the request queue and attempts to allocate it to the node with the minimum energy consumption among the available nodes. For each request, the algorithm iterates through the available node list and checks if the current node has enough resources to fulfill the request. If it does, the algorithm calculates the energy consumption of the node processing that request and compares it with the current minimum energy consumption. If the energy consumption is less than the current minimum value, the allocated node and minimum energy consumption are updated. After iterating through all the nodes for each request, the algorithm adds the allocated node to the allocation mapping allocationMap (Line 8-23). Finally, the algorithm returns the allocationMap, which indicates the mapping of each request to its assigned node (Line 24).

\begin{algorithm}[!htb]
\caption{Power-efficient Algorithm for GPT requests}
\label{Power-efficient Optimization}
\begin{algorithmic}[1]
\State \textbf{Input:} gptRequestQueue, availableNodes
\State \textbf{Output:} allocationMap

\State allocationMap $\gets \{\}$ \Comment{Initialize an empty mapping of requests to nodes}

\For{\textbf{each} request \textbf{in} gptRequestQueue}
    \State request.estimateResourceDemand() \Comment{Estimate resource demand for each request}
\EndFor
\State gptRequestQueue.sortByDescendingResourceDemand() \Comment{Sort requests by descending resource demand}

\For{\textbf{each} request in gptRequestQueue}
    \State minPower $\gets \infty$ \Comment{Initialize minimum power to positive infinity}
    \State allocatedNode $\gets \text{NULL}$ \Comment{Initialize the allocated node to NULL}

    \For{\textbf{each} node in availableNodes}
        \If{node has enough resources for request} \Comment{Check if the node has enough resources for the request}
            \State power $\gets$ estimatePower(node, request) \Comment{Estimate power consumption for allocating request to the node}
            \If{power $<$ minPower} \Comment{Update the minimum power and allocated node if a more power-efficient option is found}
                \State allocatedNode $\gets$ node
                \State minPower $\gets$ power
            \EndIf
        \EndIf
    \EndFor

    \If{allocatedNode $\neq \text{NULL}$} \Comment{Allocate the request to the most power-efficient node}
        \State allocationMap[request] $\gets$ allocatedNode
    \EndIf
\EndFor

\State \Return allocationMap \Comment{Return the allocation mapping}
\end{algorithmic}
\end{algorithm}

\section{Conclusions and Future Research Directions}\label{sec: Future Research Directions}

In this paper, we first introduced popular GPT-based models, identified the unique characteristics of resource management for GPT-based model, and discussed corresponding evaluation metrics. Based on this, we further analyzed specific challenges in resource management for GPT-based model. To achieve effective resource management, we also introduced a comprehensive resource management framework consisting of several key components. Additionally, based on our resource management framework, we proposed three scheduling algorithms specifically designed for the GPT-based model for different objectives. Finally, we explored future research directions for resource management for GPT-based model , highlighting some potential areas worth investigating. We hope that research in these areas will draw the attention of researchers and drive continuous innovation and development in resource management for GPT-based model .

Although the GPT-based model has developed rapidly and gained widespread applications in various fields, there is still significant room for improvement in resource management for GPT-based model. In the future, we need to explore more efficient and intelligent resource management techniques and optimization strategies to meet the growing resource demands of the GPT-based model. We summarize several future research directions for resource management for GPT-based model as follows:

\textbf{Specialized Hardware for GPT-based Model:} As the scale of the GPT-based model continues to increase, the computational demands also grow. Future research will focus on providing higher-performance hardware support for the GPT-based model. Manufacturers can develop specialized AI chips (such as GPUs, FPGAs, etc.) optimized for the characteristics of the GPT-based model to meet the demands of large-scale parallel computations.

\textbf{Benchmarks for Performance Evaluation:} Currently, there is a lack of standardized benchmarks  for resource management for  GPT-based model, and existing evaluation metrics are not comprehensive. Future efforts should establish more comprehensive test suites to evaluate resource management from multiple dimensions.

\textbf{Resource Utilization Maximization:} Future research needs to investigate more efficient resource management techniques to maximize the utilization of resources by the GPT-based model. Cloud data centers suffer from the issue in low resource utilization. This can be achieved through dynamic resource allocation, resource sharing, and parallel computing algorithms designed for GPT-based models.

\textbf{Scheduling Algorithms and Metrics:} To effectively perform task scheduling and resource management, future research needs to design specialized scheduling and optimization algorithms for the GPT-based model. These algorithms should consider task priorities, resource constraints, and optimization metrics to achieve better resource allocation and task scheduling strategies. Additionally, attention should be paid to scheduling algorithm metrics, such as throughput, response time, task completion rate to evaluate the effectiveness of the algorithms. Moreover, novel metrics can also be proposed to measure the metrics suitable for resource provisioning algorithms for GPR-based model.

\textbf{Security Management:} As the GPT-based model is widely used in various fields now, security concerns in resource management become increasingly prominent. Future research directions include: 1) \textit{data privacy} to ensure effective protection of user data during the GPT-based model training and inference to avoid data breaches. 2) \textit{model security} to  prevent malicious attackers from tampering with the GPT-based model to protect the integrity and reliability of model parameters and prevent degradation of model performance. 3) \textit{system security} to implement security measures to protect computational resources from malicious attacks, preventing resource abuse, occupation, or damage, and ensuring the reliability of resource management.

\bibliographystyle{plain}
\bibliography{1}

\end{document}